\documentclass[conference]{IEEEtran}
\usepackage{amsmath,amssymb,amsfonts}
\usepackage{algorithmic}
\usepackage{graphicx}
\usepackage{textcomp}
\usepackage{xcolor}
\usepackage[utf8]{inputenc}
\usepackage[T1]{fontenc}
\usepackage{amsmath}
\usepackage{subfig}
\usepackage{authblk}
\usepackage{xcolor}
\usepackage[noadjust]{cite}
\usepackage[font={footnotesize}]{caption}
\usepackage{url}
\usepackage{comment}
\usepackage{balance}

\newcommand{\sprac}{S-PRAC}

\def\BibTeX{{\rm B\kern-.05em{\sc i\kern-.025em b}\kern-.08em
		T\kern-.1667em\lower.7ex\hbox{E}\kern-.125emX}}

\title{S-PRAC: Fast Partial Packet Recovery with Network Coding in Very Noisy Wireless Channels}

\author[1,2]{Kurniawan D. Irianto}
\author[1]{Juan A. Cabrera}
\author[1]{Giang T. Nguyen}
\author[1]{Hani Salah}
\author[1]{Frank H. P. Fitzek}
\affil[1]{Deutsche Telekom Chair of Communication Networks, TU Dresden, Germany}
\affil[2]{Department of Informatics, Universitas Islam Indonesia, 55584 Yogyakarta, Indonesia}
\affil[ ]{E-mail: \{kurniawan\_dwi.irianto@tu-dresden.de, k.d.irianto@uii.ac.id\}, \{juan.cabrera \textbar \ giang.nguyen \textbar \ \protect\\ hani.salah \textbar \ frank.fitzek\}@tu-dresden.de}

\begin{document}

\maketitle

\begin{abstract}
Well-known error detection and correction solutions in wireless communications are slow or incur high transmission overhead. Recently, notable solutions like PRAC and DAPRAC, implementing partial packet recovery with network coding, could address these problems. However, they perform slowly when there are many errors. We propose S-PRAC, a fast scheme for partial packet recovery, particularly designed for very noisy wireless channels. S-PRAC improves on DAPRAC. It divides each packet into segments consisting of a fixed number of small RLNC-encoded symbols, and then attaches a CRC code to each segment and one to each coded packet. Extensive simulations show that S-PRAC can detect and correct errors quickly. It also outperforms DAPRAC significantly when the number of errors is high.\\
\end{abstract}

\section{Introduction} \label{sec:introduction}

Errors are inevitable in wireless communications. They can be attributed, for instance, to signal fluctuation, fading, or interference. Nevertheless, applications usually require a certain level of reliability for data transmission. Audio and video applications can only tolerate some errors to recover without significant distortion. Data applications even require error-free transmission.

To detect errors, parity check and Cyclic Redundancy Check (CRC) schemes have been proposed and implemented extensively. To  tackle the bit error issue, well-known solutions like Forward Error Correction (FEC) \cite{fec_costello1982error} and Automatic Repeat Request (ARQ) \cite{arq_1091865} have been proposed. Approaches based on FEC add significant redundant information into transmitted packets, so that the original packets can be recovered even with the presence of some corrupted bits. In a different approach, ARQ explicitly triggers a retransmission of a packet when an error is detected inside the packet. However, both of the above approaches are inefficient when only a few bits of the whole data block are corrupted. In particular, FEC introduces a significant overhead to the transferred payload, while ARQ introduces latency due to retransmission.

Packetized Rateless Algebraic Consistency (PRAC) \cite{Angelopoulos2014} and Data Aware PRAC (DAPRAC) \cite{Cabrera2017} are notable solutions for the above issues. They use a combination of an Algebraic Consistency Rule (ACR) and CRC checks at the decoder side to recover corrupted bits. Subsequently, they reduce the overhead of FEC and latency due to retransmission in ARQ.  Specifically, for encoding, they employ a linear cross-packet code over a block of $k$ packets. Furthermore, they divide each packet into $l$ symbols. Such a block of packets and symbols is considered as a matrix of size $k \times l$. The coded packets are the result of multiplying the matrix with a randomly generated matrix of coefficients. As a result, the receiver only needs $x$ out of $k$ encoded packets to decode and find original packets. More importantly, this formulation allows to apply ACR, meaning that linear combinations of the $x$ packets can be use to express the remaining $(k - x)$, to detect corrupted symbols. 

Correction in PRAC and DAPRAC is an iterative process, following ACR checks. A search algorithm is used to identify the correct symbols until the ACR check for the specific column is satisfied. By the end of each iteration, the CRC code is updated. The main drawback of PRAC and DAPRAC is the prolonging search process which is required for testing all combinations of symbols to correct corrupted packets. This results in a very slow partial packet recovery, especially when the number of errors is high.

Our main contribution in this paper is Segmented Packetized Rateless Algebraic Consistency (S-PRAC), a new scheme for partial packet recovery with network coding. S-PRAC is effective and fast in very noisy channel conditions. Whereas in the previous works, such as PRAC \cite{Angelopoulos2014} and DAPRAC \cite{Cabrera2017}, the performance tends to decrease when the number of errors increase in noisy channels.

In fact, our main goal is to reduce the total time required for detecting and correcting errors. Following the divide-and-conquer approach, \sprac ~strives to identify corrupted  symbols. The key idea is twofold: First, S-PRAC divides the packets into segments, each consists of a fixed number of encoded symbols. Second, S-PRAC adds two CRC codes: (1)  CRC-32 code attached to the end of the coded packet (i.e. outer CRC) like in DAPRAC and (2) CRC-8 code attached to the end of each segment (i.e. inner CRC). This design helps to significantly lower error detection and correction times. Specifically, using a small number of symbols inside each segment, S-PRAC can reduce the number of permutations in the correction process. In addition, using the inner CRC codes can accelerate the detection and correction of errors because both processes are performed on the segment level.


\begin{figure*}[!ht]
	\centering
	\subfloat[ACR: step 1]{\includegraphics[width=1.7in]{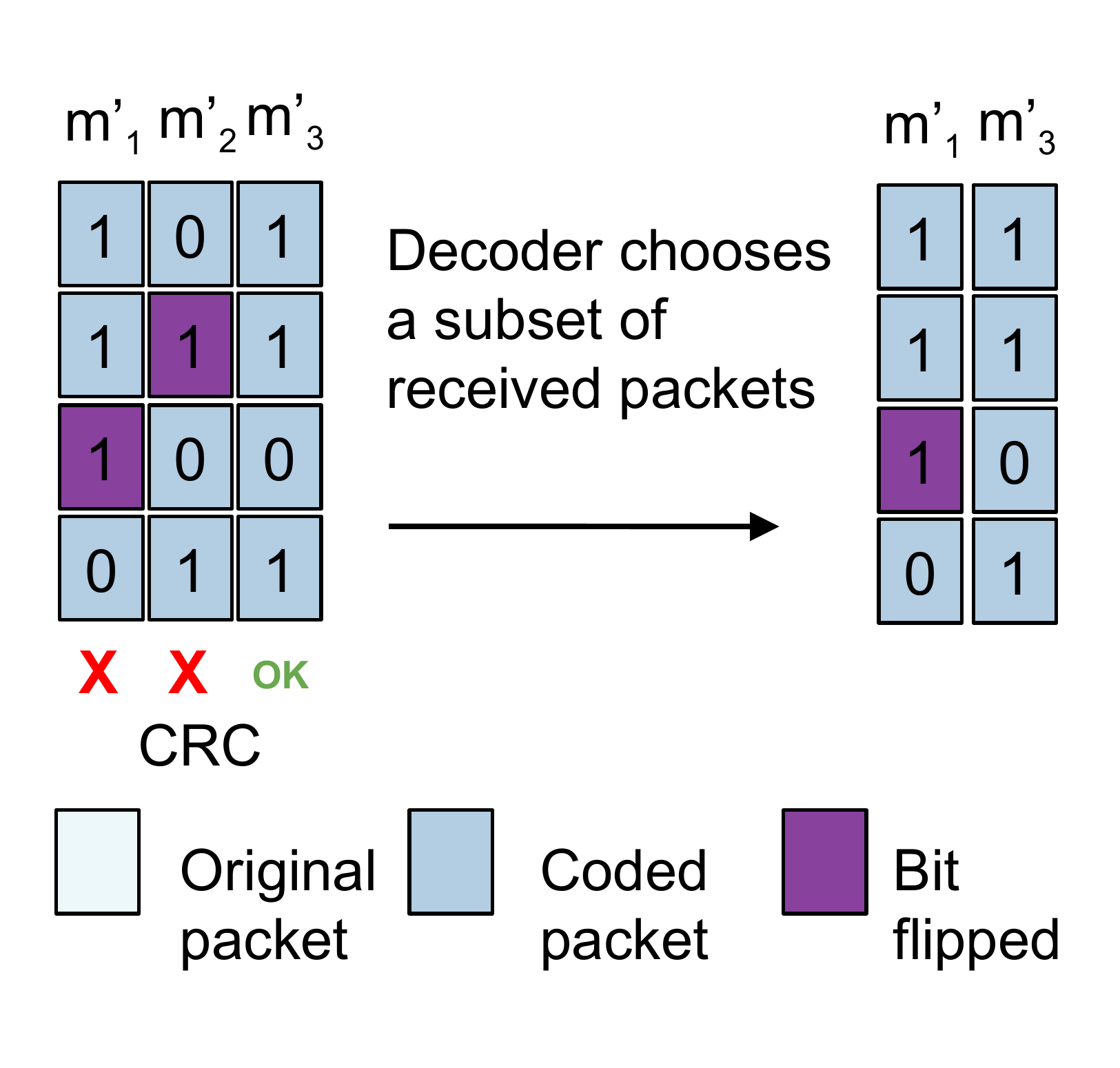}\label{fig:acr_1}}
	\hfil
	\subfloat[ACR: step 2]{\includegraphics[width=1.7in]{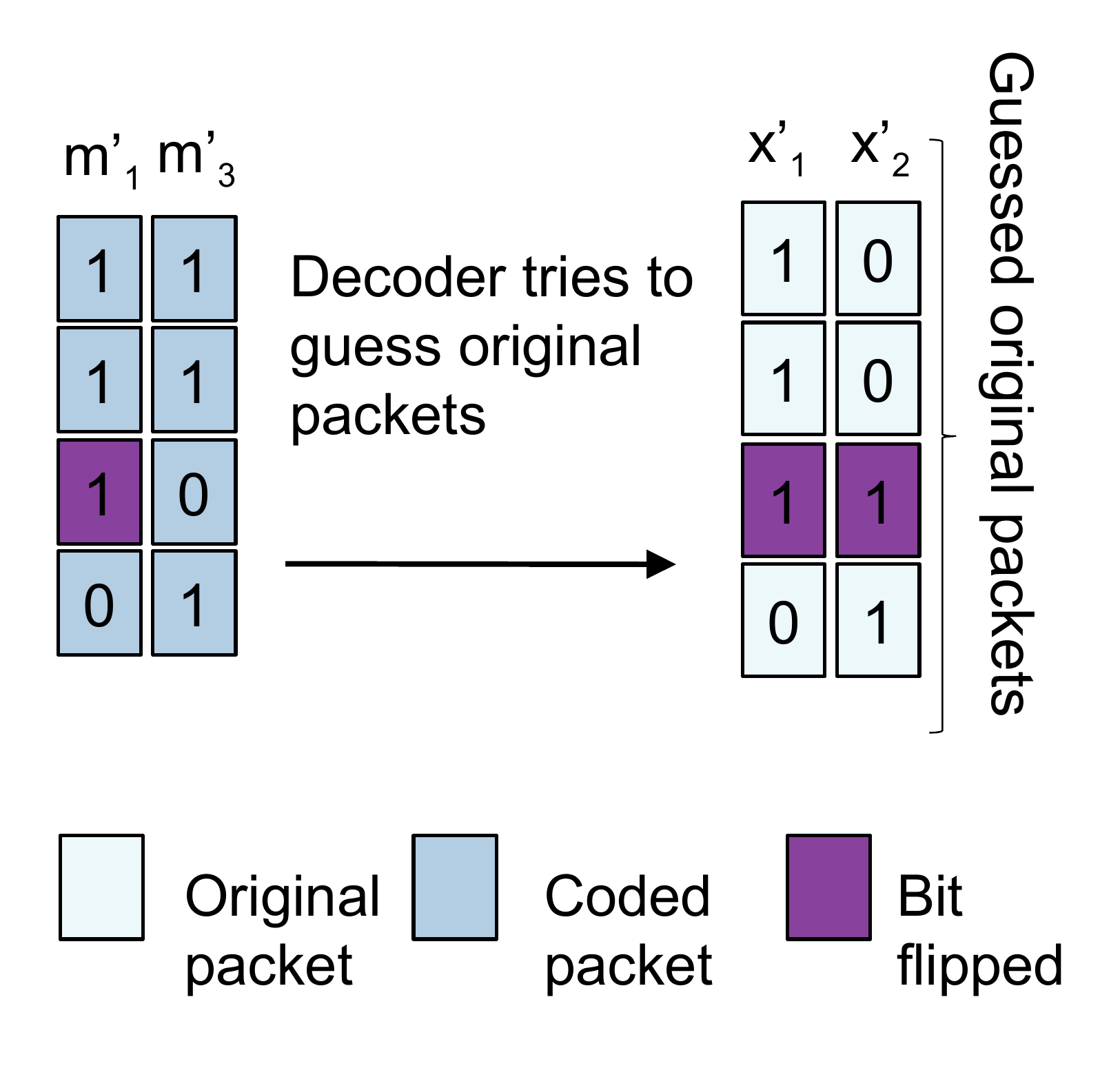}\label{fig:acr_2}} 
	\hfil
	\subfloat[ACR: step 3]{\includegraphics[width=1.7in]{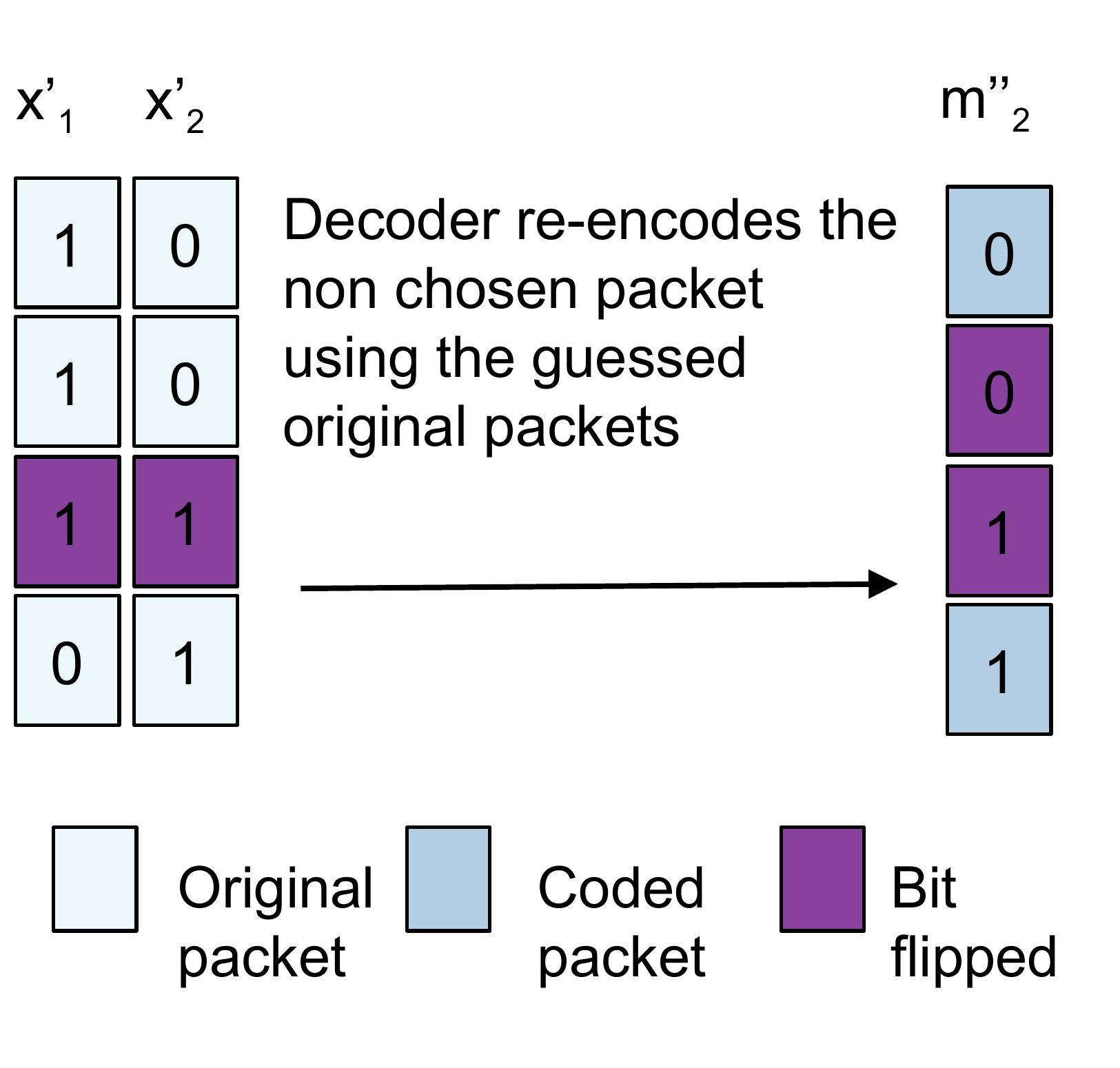}\label{fig:acr_3}}
	\hfil
	\subfloat[ACR: step 4]{\includegraphics[width=1.7in]{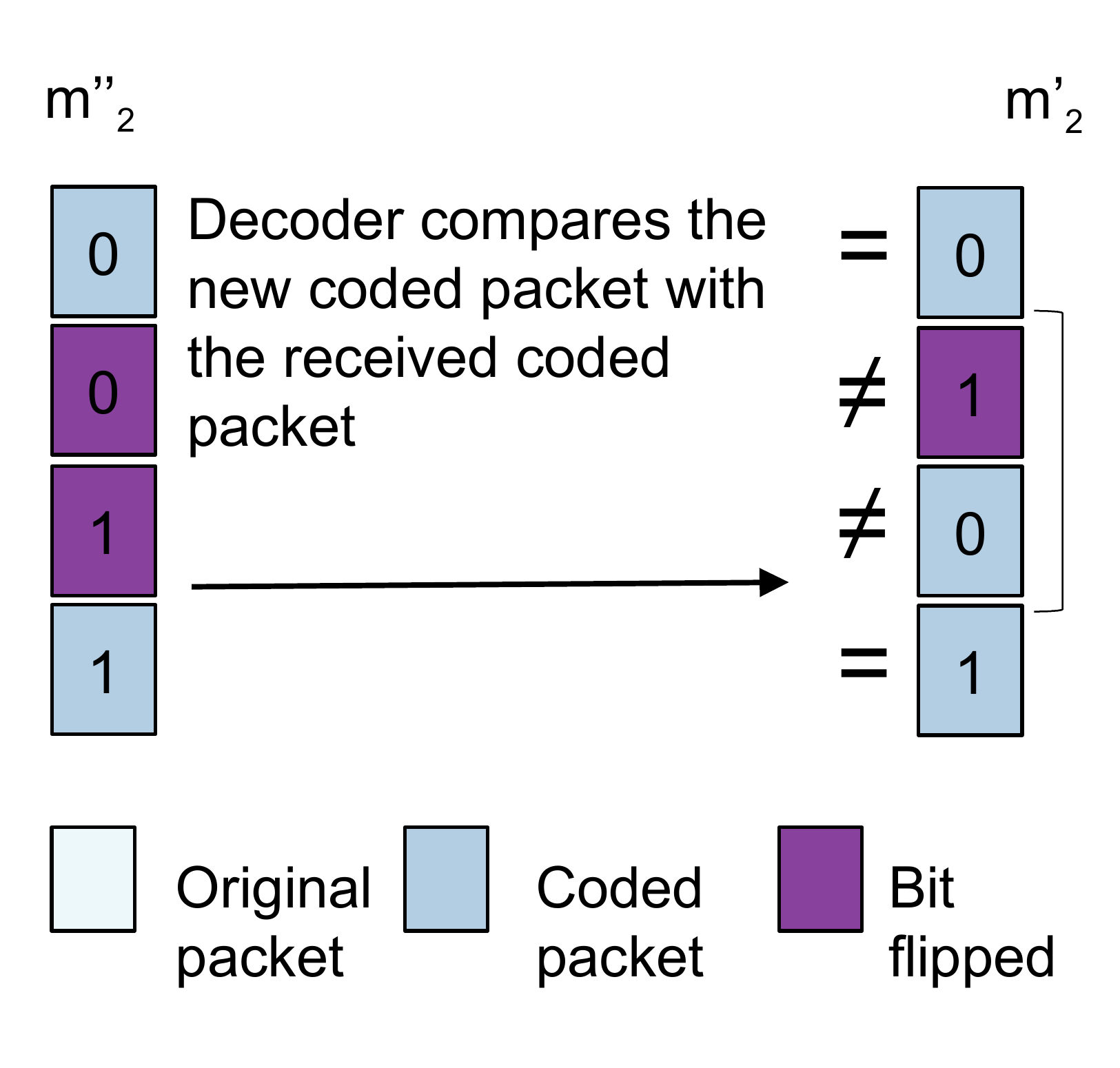}\label{fig:acr_4}}
	\caption{Steps in algebraic consistency rule (ACR) check}\label{fig:acr_steps}
\end{figure*}


The rest of the paper is organized as follows: we discuss the background and related work in Section \ref{sec:background} and elaborate the design of S-PRAC in Section \ref{sec:s-prac}. Next, Section \ref{sec:evaluation} describes our evaluation setup and results. Lastly, we conclude in Section \ref{sec:conclusion}.

\section{Background and Related Work}\label{sec:background}
In this section, firstly, we describe the background of algebraic consistency rule (ACR) check in Subsection \ref{subsec:background}. ACR check plays an important role for detecting error locations in this paper. Secondly, in Subsection \ref{sebsec:related_work}, we discuss the most notable literature on partial packet recovery (PPR). 

\subsection{Overview of ACR check}\label{subsec:background}

Algebraic consistency rule (ACR) check in network coding is, firstly, proposed in \cite{Angelopoulos2014}. The authors utilize the consistency of coded packets with network coding to estimate the error bit locations. When a sender wants to send data, it will create encoded packets from the original data and send them through the channel.

Assuming that $x\textsubscript{i}$ is original packet, $m\textsubscript{i}$ is coded packet at the sender and $m'\textsubscript{i}$ is received coded packet at the receiver, where $i\geq1$ . The receiver will receive $m'\textsubscript{i}$ and try decoding it to retrieve $x\textsubscript{i}$. Let  $x'\textsubscript{i}$ be the retrieved original packet after decoding using $m'\textsubscript{i}$ and $m''\textsubscript{i}$ be re-encoded packet after encoding using $x'\textsubscript{i}$ at the receiver. In network coding, encoded and re-encoded packets should be identical. Therefore, if the packet transmissions are received without errors, then $m'\textsubscript{i}$ and $m''\textsubscript{i}$ are identical. In this case, a packet is said to be consistent when $x\textsubscript{i}=x'\textsubscript{i}$ or $m\textsubscript{i} = m'\textsubscript{i} = m''\textsubscript{i}$. In other words, if the packets are not consistent, there should be errors in the packets. With this consistency rule check, we could identify the error locations in the partial packets.

Fig. \ref{fig:acr_steps} shows an illustrative example of ACR. Assuming the sender wants to send two original packets (i.e. $x\textsubscript{1}$ and $x\textsubscript{2}$) and creates three encoded packets (i.e. $m\textsubscript{1}$, $m\textsubscript{2}$, and $m\textsubscript{3}$). Then the receiver will receive three encoded packets, namely $m'\textsubscript{1}$, $m'\textsubscript{2}$, and $m'\textsubscript{3}$, where the first and second packets contain an error bit and false CRC code. However, the receiver does not know the location of errors. It only can detect the errors through CRC check. Therefore, in order to estimate the error locations, the receiver needs to run ACR check, and later correct the errors with permutation rule in error correction process. 

As shown in Fig. \ref{fig:acr_steps}, ACR check consists of the following four steps: 
\begin{enumerate}
	\item Choose a subset of received encoded packets: here, the decoder selects $m'\textsubscript{1}$ and $m'\textsubscript{3}$. The selected packets should consist of valid and invalid packets. It is preferable to have more valid packets than invalid packets.
	\item Guess the original packets: decoder tries to retrieve the original packets (i.e. $x'\textsubscript{1}$ and $x'\textsubscript{2}$) by decoding using random linier network coding (RLNC). 
	\item Re-encode the non chosen packet: by using the guessed original packets, the decoder needs to re-encode the non chosen packet (i.e. $m''\textsubscript{2}$).
	\item Compare the re-encoded packets with the received encoded packets: lastly, the decoder estimates the error locations of $m'\textsubscript{2}$ by comparing $m'\textsubscript{2}$ and $m''\textsubscript{2}$, where the different bits point out the location of errors.
\end{enumerate}

\subsection{Related Work}\label{sebsec:related_work}
To the best of our knowledge, PPR was firstly proposed in \cite{Sindhu1977}. The author suggests to buffer packets received with errors, instead of discarding them, since they may contain useful information. If the retransmission of the same packet also contains errors, the joint information present in   the first packet and the retransmission can be used to correct errors in the two packets. If errors remain, further retransmissions are performed until enough redundancy is present for correction. In \cite{Miu2005}, PPR is proposed for IEEE 802.11 WLANs. In particular, the end device performs PPR on multiple copies of the same packet received from different access points. A similar approach, called ZipTx, is proposed in \cite{Lin2008}. It uses error correction codes and known pilot bits in each packet when bit error rates (BER) are low and high, respectively. ZipTx recovers corrupted packets using retransmission without coding.

Some PPR approaches exploit physical layer information. For instance, the authors of \cite{Jamieson2007} propose two PPR approaches: (1) SoftPHY which provides hints from physical layer interface to higher layers and (2) a postamble scheme for recovering packets with corrupted preambles.  Jamieson and Balakrishnan \cite{Jamieson2007} design a link layer protocol named PP-ARQ. In this protocol, the receiver specifies in the acknowledgment the parts of the packet that are received without errors. The transmitter, in turn, retransmits only the parts of the packet that are not listed in the acknowledgment, i.e. the corrupted parts. MIXIT \cite{Katti2008} employs two methods for improving the throughput in wireless mesh networks: (1) increased concurrency and (2) congestion-aware forwarding. In SOFT \cite{Woo2007}, which aims to handle dead spots and high loss rates, the physical layer transfers its confidence in bits to the higher layers. A recent solution, called CodeRepair, is proposed by Huang et al \cite{Huang2015}. CodeRepair uses the single parity code (SPC) for correcting bit errors in packets. SPC offers a simple error detection code. In addition, CodeRepair integrates several techniques, such as code reversing, selective re-decoding, and parity sampling  rate optimization, to utilize SPC for a high efficient FEC in correcting errors.

Other works propose PPR approaches based on bit rate adaptation. For instance, SampleRate \cite{Bicket2005}, aiming to optimize the throughput on wireless links, adjusts the bit rate according to estimation of per-packet transmission time. Each bit rate uses a particular modulation scheme 
to transform a data stream into a series of encoded symbols. Vutukuru et al. \cite{Vutukuru2009} propose SoftRate, a protocol that is highly resposinve to channel conditions. More precisely, SoftRate uses both bit rate adaptation and physical layer information. It estimates channel BER using confidence information calculated from the physical layer, and passes it to higher layers. Implementation and measurement of a cross-layer framework for rate adaptation in urban and vehicular environments are presented in \cite{Camp2010}. The study considers fast-fading, multi-path, and interference for measuring rate adaptation accuracy in diverse channel conditions.

Other PPR approaches, commonly called segment-based PPR, divide the packet into segments, and retransmit only the corrupted segments. Notable examples include \cite{Ganti2006}, \cite{Iyer2009}, and \cite{Han2010}. Other approaches \cite{Angelopoulos2014, Cabrera2017, Mohammadi2015a, Mohammadi2016} exploit the properties of network coding. The approach presented in \cite{Mohammadi2015a} and \cite{Mohammadi2016} is based on random linear network coding (RLNC), sparse error recovery, and compressive sensing. It exploits partial packets by solving a set of standard sparse recovery problems for error estimation and correction. PRAC \cite{Angelopoulos2014, Angelopoulos2017} performs error detection and correction on partial packets by applying multiple rounds of algebraic consistency rule (ACR) checks. It does not use physical layer information, and requires minimum feedback for notification of completion. 
Error detection and correction times in PRAC are slow, and they are not affected by the number of errors. DAPRAC \cite{Cabrera2017} is a modification of PRAC that aims to reduce the time of the decoding process. The solution that we present in this paper improves on DAPRAC. Given this relevance of DAPRAC to our work, we provide a detailed overview below. 

DAPRAC implements a column-based approach for locating errors, where each column represents a symbol over a finite field. It performs an ACR iteration to  identify corrupted bits in each column. It corrects the errors using a brute-force method with permutation of corrupted bits. More precisely, once corrupted bits are located, DAPRAC counts the number of possible permutations of these bits, and matches each permutation with CRC values. The incorrect CRC values are then updated with the correct ones. After correcting the corrupted bits, the decoder decodes the coded packets, using RLNC, to retrieve the original information. Although this design lowers the decoding time achieved with PRAC, it still results in slow decoding in very noisy channels, i.e. when the number of errors is high. This is because the more the errors, the higher the number of needed permutations.

Our proposed approach for PPR with network coding, named S-PRAC, is different from the previous works. In particular, we utilize the segmentation over packets to make error detection and correction faster. Moreover, we insert a CRC code to each segment so that S-PRAC can only focus on the corrupted segments. As a result, the needed time for detection and correction is remarkably reduced. Therefore, our scheme is faster than prior solutions when the error numbers are high, especially in noisy channel environments.

\section{Partial Packet Recovery with S-PRAC} \label{sec:s-prac}

We describe in this section S-PRAC, our solution for partial packet recovery. We present the design concepts of S-PRAC in Subsection \ref{idea}. After that, we detail how packets are encoded and decoded in Subsection \ref{enc} and Subsection \ref{dec}, respectively. 


\subsection{Design Concepts} \label{idea}

\begin{figure}[!t]
	\centering
	\includegraphics[width=\linewidth]{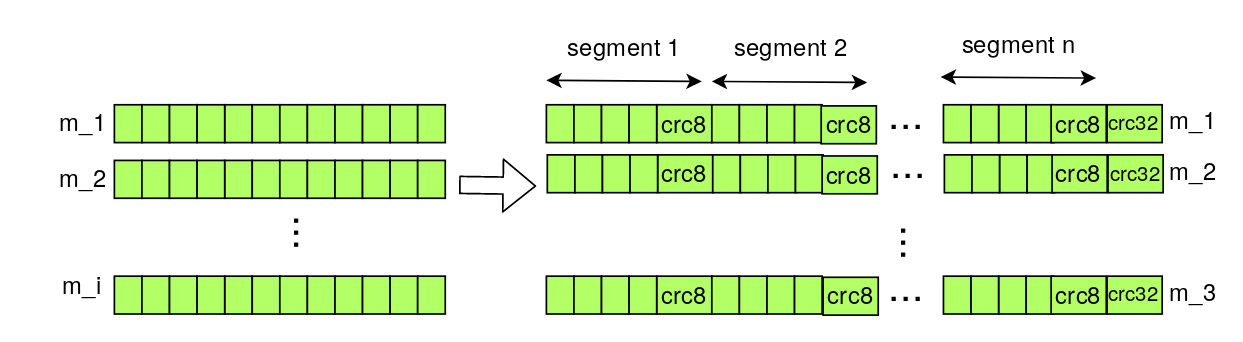}
	\caption{S-PRAC concepts: segmentation of coded packets, adding an inner CRC-8 code to each segment, and adding an outer CRC-32 code to each coded packet after segmentation.} 
	\label{fig:comm_sys&encoder}
\end{figure}

Channel noise can cause bits flipping. A packet with one or more flipped bits is called corrupted packet or invalid packet. S-PRAC aims to detect and correct corrupted packets quickly in very noisy wireless channels. Towards that end, it performs RLNC-based encoding and decoding on partial packets. More precisely, as can be seen in Fig. \ref{fig:comm_sys&encoder}
, the transmitter divides each packet into equal-sized segments. Each segment contains symbols encoded with network codes, over a Galois field 
 \cite{Sun2015}. The transmitter also adds an outer CRC-32 code at the end of the segmented packet (like PRAC \cite{Angelopoulos2014} and DAPRAC \cite{Cabrera2017}) and, additionally, an inner CRC-8 code at the end of each segment. 

\begin{figure}[!t]
	\centering
	\begin{tabular}{@{}c@{}}
		\includegraphics[width=\linewidth]{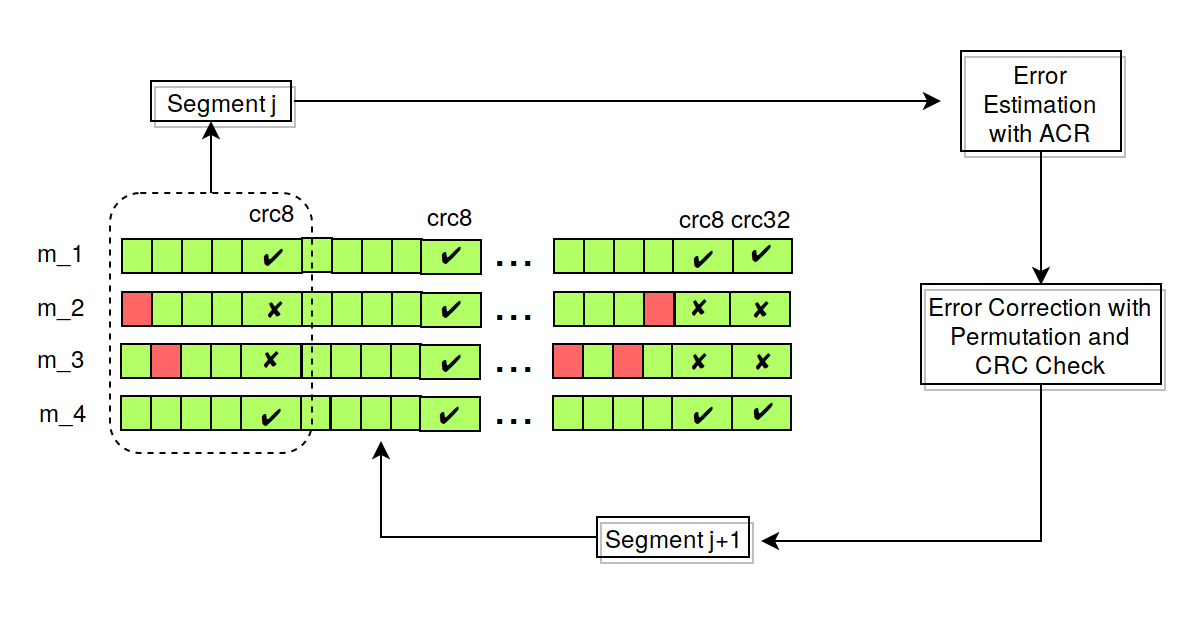} \\[\abovecaptionskip]
		\small (a)
	\end{tabular}
	
	\vspace{\floatsep}
	
	\begin{tabular}{@{}c@{}}
		\includegraphics[width=\linewidth]{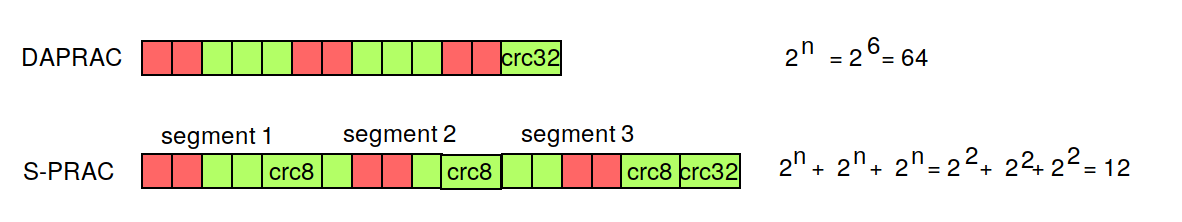} \\[\abovecaptionskip]
		\small (b)
	\end{tabular}
	
	\caption{(a) S-PRAC's error estimation and error correction processes run at the decoder side for each segment; (b) The number of permutations in S-PRAC and DAPRAC (in this example, number of corrupted bits is 6, and number of segments in S-PRAC is 3).}\label{fig:sprac&perm}
\end{figure}

Recovery of segments in S-PRAC, at the receiver side, is illustrated in Fig. \ref{fig:sprac&perm}a. The receiver cannot determine the exact locations of corrupted bits. It rather estimates these locations by applying multiple rounds of ACR checks. The correction process is triggered immediately after each ACR round, before starting the next ACR round. In the correction process, the receiver performs a brute-force search for the correct permutation of corrupted bits, i.e. the permutation that matches the segment's inner CRC. The more the corrupted bits, the more the number of permutations, thus the longer the time needed to find the right permutation. Segmentation can reduce the correction time by limiting the number of permutations required for correction and by distributing the errors over the segments. S-PRAC performs error estimation and correction on the segments one by one, in order. After the last segment, the outer CRC code is updated. Lastly, the corrected packet is decoded with RLNC.

The number of permutations ($N$) required for correction depends both on the symbol's field size ($q$) as well as on the number of corrupted bits ($n$). 
 $N$ can be calculated in DAPRAC as

\begin{equation}\label{eq: permu_prac}
N\textsubscript{DAPRAC} = q\textsuperscript{n}
\end{equation} while in S-PRAC as
\begin{equation}\label{eq: perm_sprac}
N\textsubscript{S-PRAC} = \sum_{i=1}^{i=k} q\textsubscript{i}^{n}
\end{equation} where $i$ ($1 \leqslant i \leqslant k$) is number of segments. 

\vspace{5pt}
The example in Fig. \ref{fig:sprac&perm}b compares the number of permutations ($N$) in DAPRAC and S-PRAC. Applying Eq. \ref{eq: permu_prac} and Eq. \ref{eq: perm_sprac}, using $q = 2$ and $n = 6$, will result in $N = 64$ in DAPRAC and $N = 12$ in S-PRAC. That is to say, the correction is more than five times faster with S-PRAC than with DAPRAC, in this example.


\subsection{Encoding Process} \label{enc}

We assume that there are $K$ original packets $M\textsubscript{1}, M\textsubscript{2}, ..., M\textsubscript{K}$ produced by one transmitter and need to be sent to one receiver. Each packet is T bits long, i.e. $M\textsubscript{j} = \{m\textsubscript{j1},m\textsubscript{j2},...,m\textsubscript{jT}\}$. 
The encoder transforms the $K$ original packets into $N$ coded packets $X\textsubscript{1}, X\textsubscript{2}, ..., X\textsubscript{N}$, where $N > K$. Each coded packet $X\textsubscript{i}= \{x\textsubscript{i1}, x\textsubscript{i2}, ..., x\textsubscript{iT}\}$ is related to a series of randomly selected coefficients $G\textsubscript{i}=\{g\textsubscript{i1}, g\textsubscript{i2}, ..., g\textsubscript{iK}$\}. It creates a linear combination of the original packets, and is equivalent to
\begin{equation} 
\label{eq:1} x\textsubscript{it} = \sum_{j=1}^{K} g\textsubscript{ij} \times m\textsubscript{jt},
\end{equation} 
where $1\leq t \leq T$ and $1 \leq i \leq N$. We assume that the coded packet has the coefficients $g$, a named coding coefficient, and the encoded data $x$ (Eq. \ref{eq:1}). In matrix notation, the encoding process can be  defined as 
\begin{equation}
X = G \times M
\end{equation} where $M$ is the matrix of the original packets, $G$ is the matrix of the coding coefficients, and $X$ is the matrix of the coded packets.

Segmentation is performed after encoding the packets. Then, one inner CRC-8 code is inserted to each segment, and one outer CRC-32 code is inserted to each encoded packet. While increasing the number of segments will increase the number of inner CRC codes, thus computations, it can significantly decrease the correction time. 


\subsection{Decoding Process} \label{dec}

The decoding process consists of two subsequent phases: (1) the estimation and correction phase and (2) the decoding phase. In general, the decoder receives the coded packets, checks their outer CRC values, and classifies them accordingly into valid packets and invalid packets. The packets then stay in a buffer awaiting for the estimation and correction phase. The decoder starts the first phase after receiving $g+1$ packets, where $g$ is the number of symbols (i.e. generation size in RLNC). The decoding phase is initiated directly after the first phase. 

Before starting correction of partial packets in the first phase, we need to estimate error locations using Algebraic Consistency Rule (ACR) checks, as proposed in \cite{Angelopoulos2014, Angelopoulos2017}. Essentially, the ACR mechanism consists of four steps. First, it selects a subset of received coded packets, containing  the valid and invalid packets. Second, it reconstructs the uncoded packets by guessing them. Third, using the guessed uncoded packets, ACR re-encodes the non chosen packet. Fourth and last, it compares the re-encoded packets with the received encoded packets. The details of ACR check are already explained in subsection \ref{subsec:background}. 

ACR is performed segment by segment over the matrix of received coded packets. Therefore, multiple ACR rounds are needed to find all corrupted bits in all segments. The decoding phase is initiated after the successful completion of the estimation and correcting phase. In this phase, the coded packets are decoded using RLNC.



\begin{figure*}[!t]
	\centering
	\subfloat[Error estimation delay]{\includegraphics[width=3.5in]{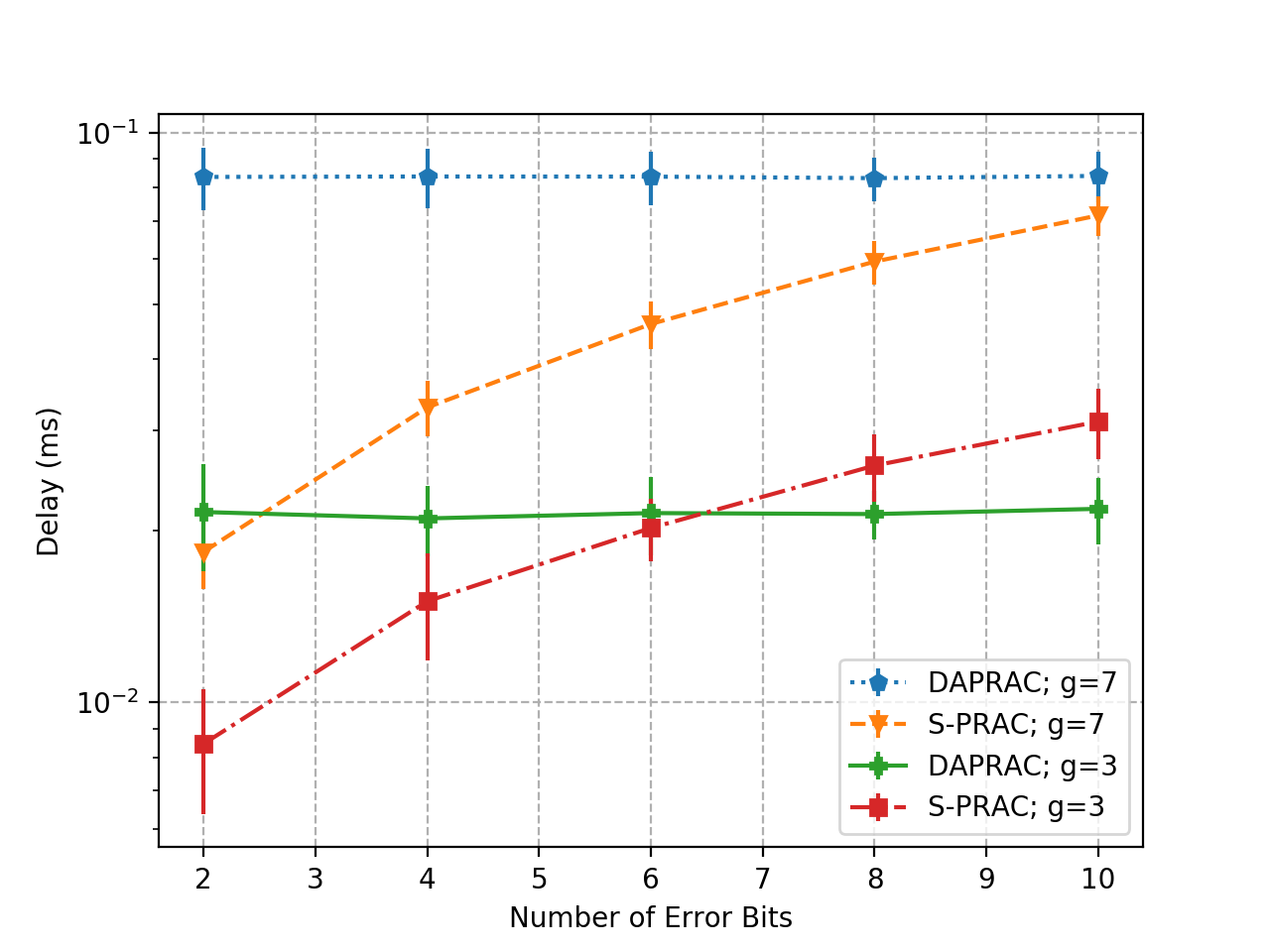}
		\label{fig:err_estimation}}
	\hfil
	\subfloat[Error correction delay]{\includegraphics[width=3.5in]{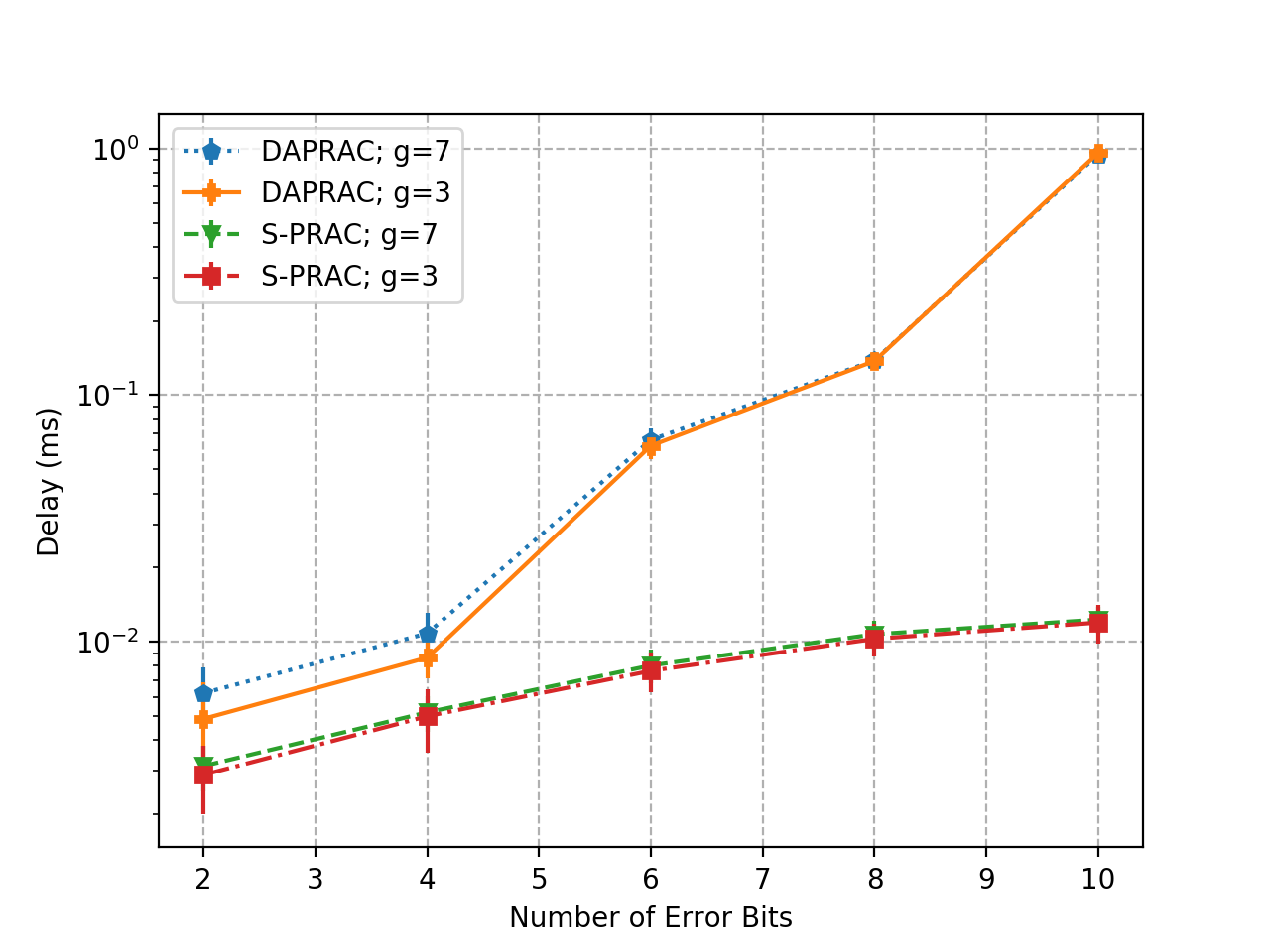}
		\label{fig:err_correction}}
	\caption{Error estimation delay and error correction delay (S-PRAC vs. DAPRAC)}
	\label{fig:estimation_correction}
\end{figure*}


\begin{figure*}[!t]
	\centering
	\subfloat[Impact of the number of corrupted bits (S-PRAC vs. DAPRAC)]{\includegraphics[width=3.5in]{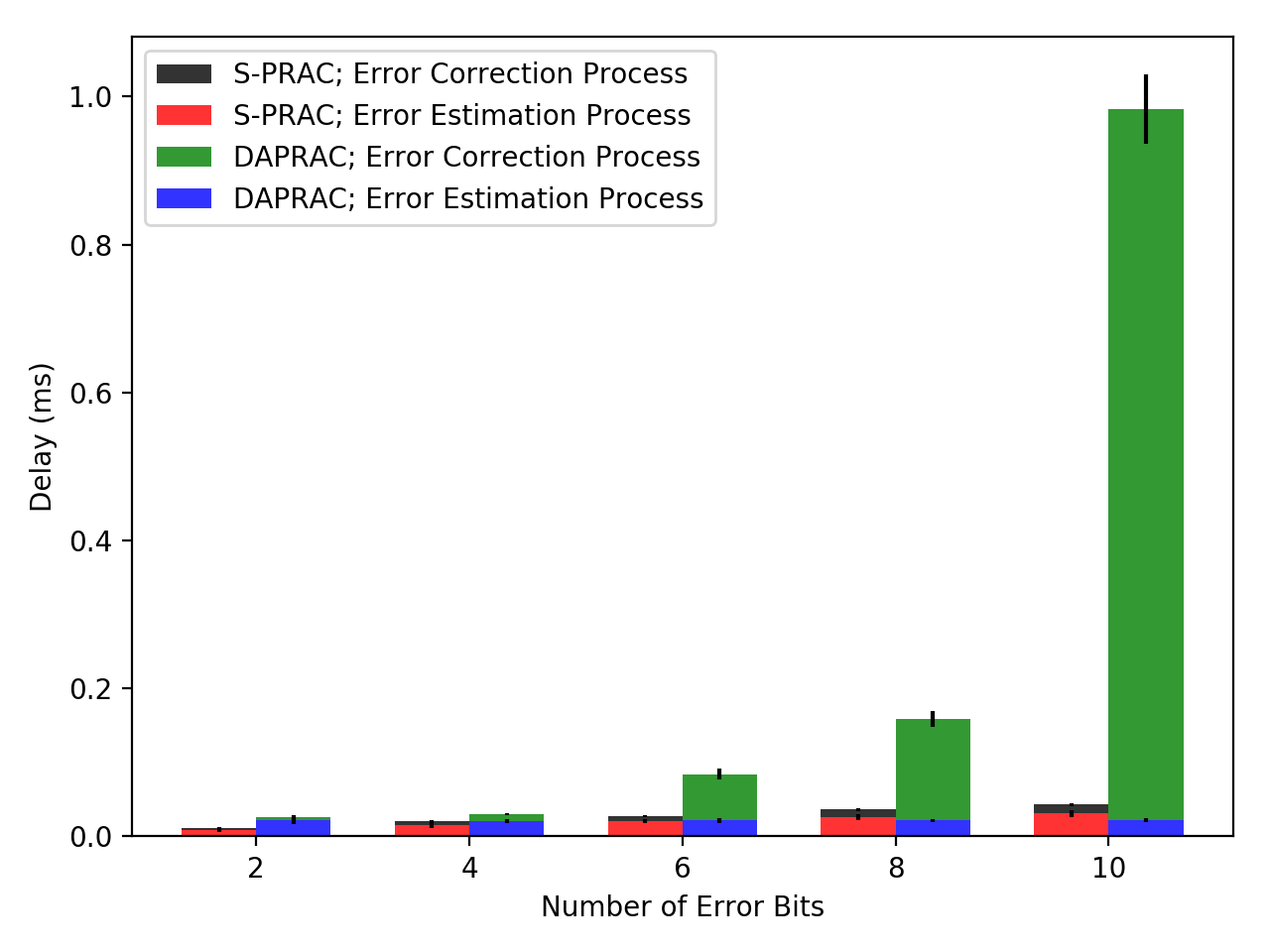}
		\label{fig:estimation_corr}}
	\hfil
	\subfloat[Impact of the number of segments in S-PRAC]{\includegraphics[width=3.5in]{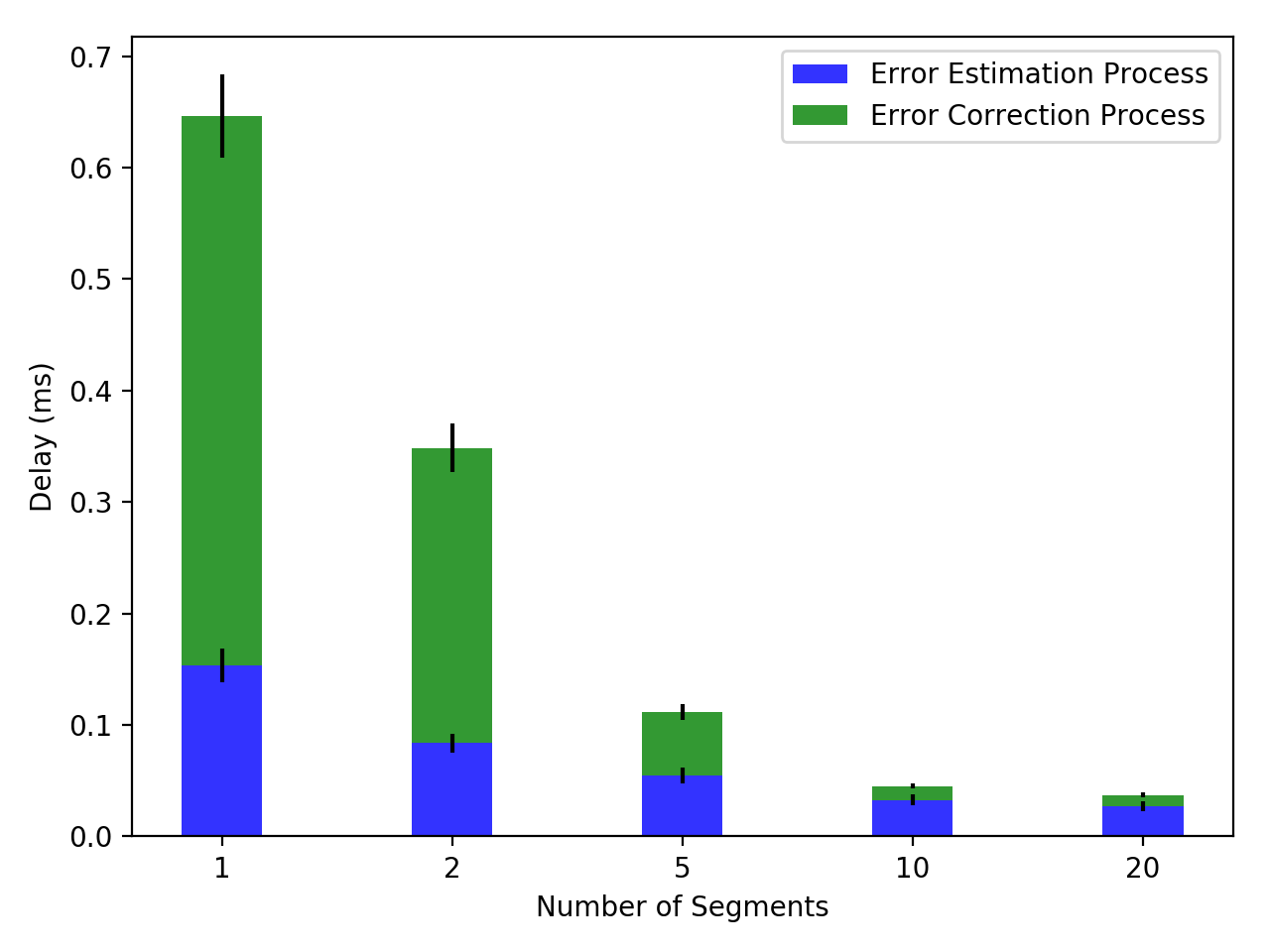}
		\label{fig:segmentation}}
	\caption{Total of error estimation delay and error correction delay}
	\label{fig:est_corr_segmentation}
\end{figure*}


\begin{figure*}[!t]
	\centering
	\subfloat[Encoding delay]{\includegraphics[width=3.5in]{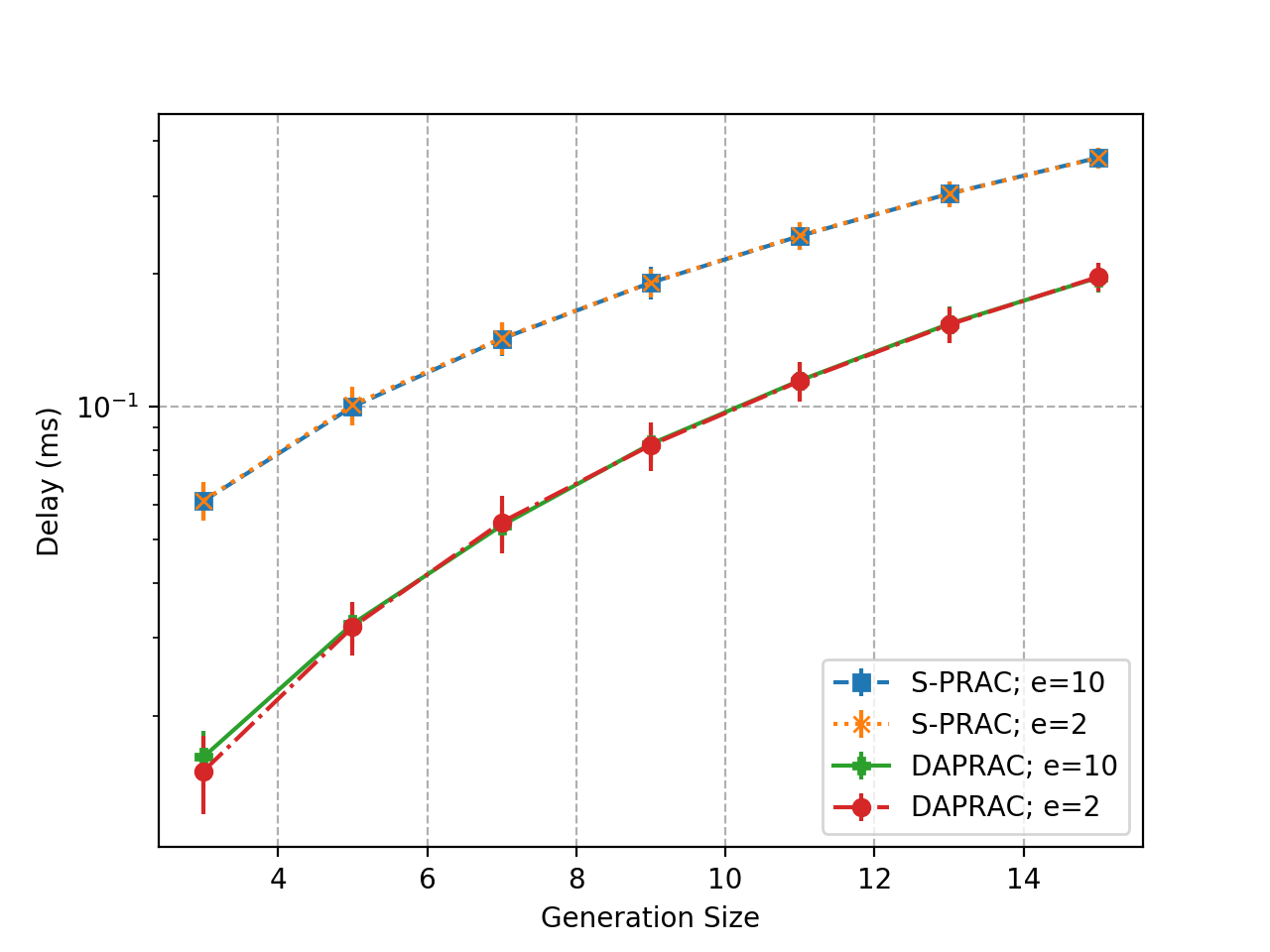}
		\label{fig:encoding}}
	\hfil
	\subfloat[Decoding delay]{\includegraphics[width=3.5in]{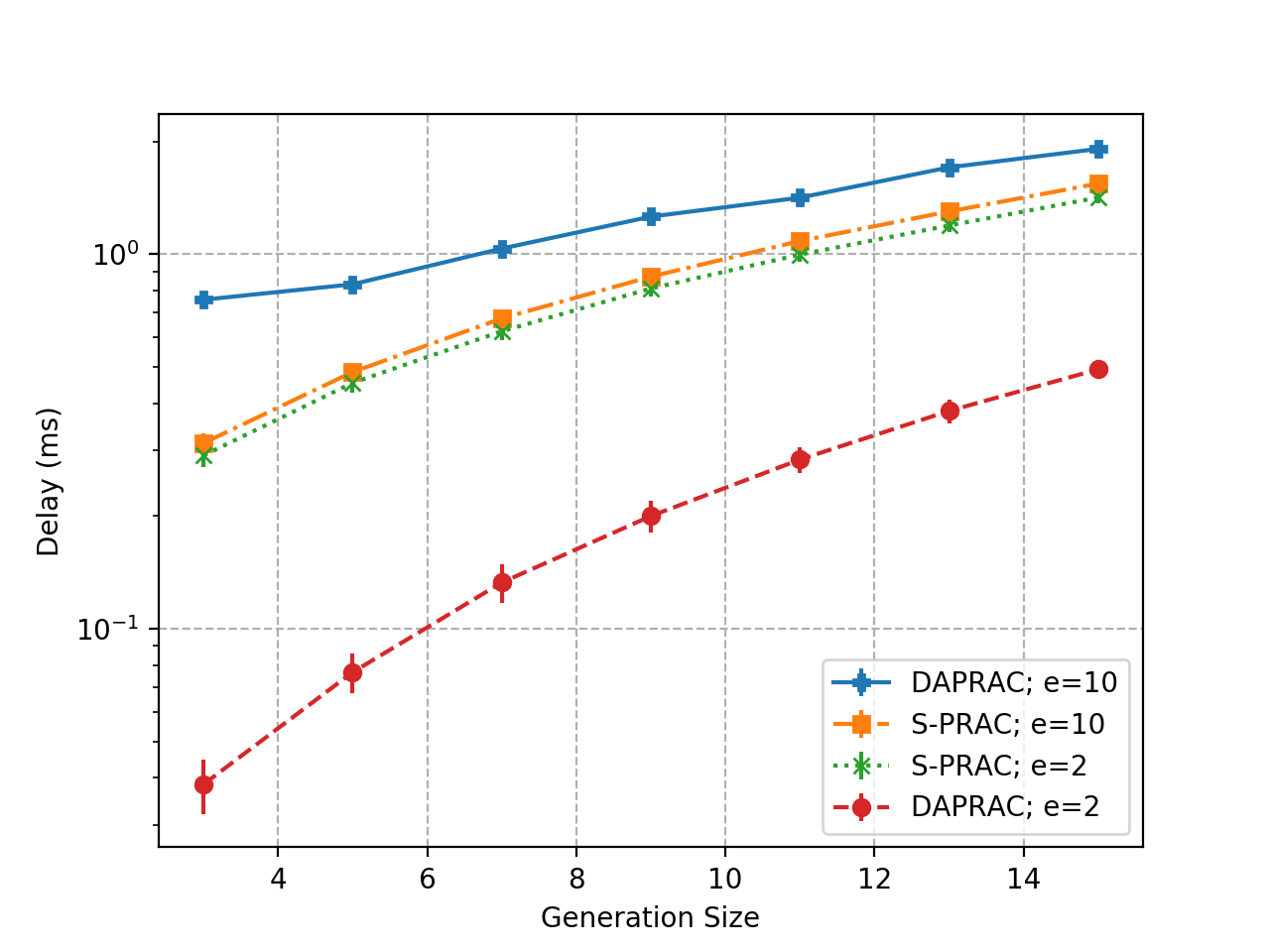}
		\label{fig:decoding}}
	\caption{Encoding delay and decoding delay (S-PRAC vs. DAPRAC)}
	\label{fig:encoding_decoding}
\end{figure*}


\begin{figure}[!t]
	\centering
	\includegraphics[width=\linewidth]{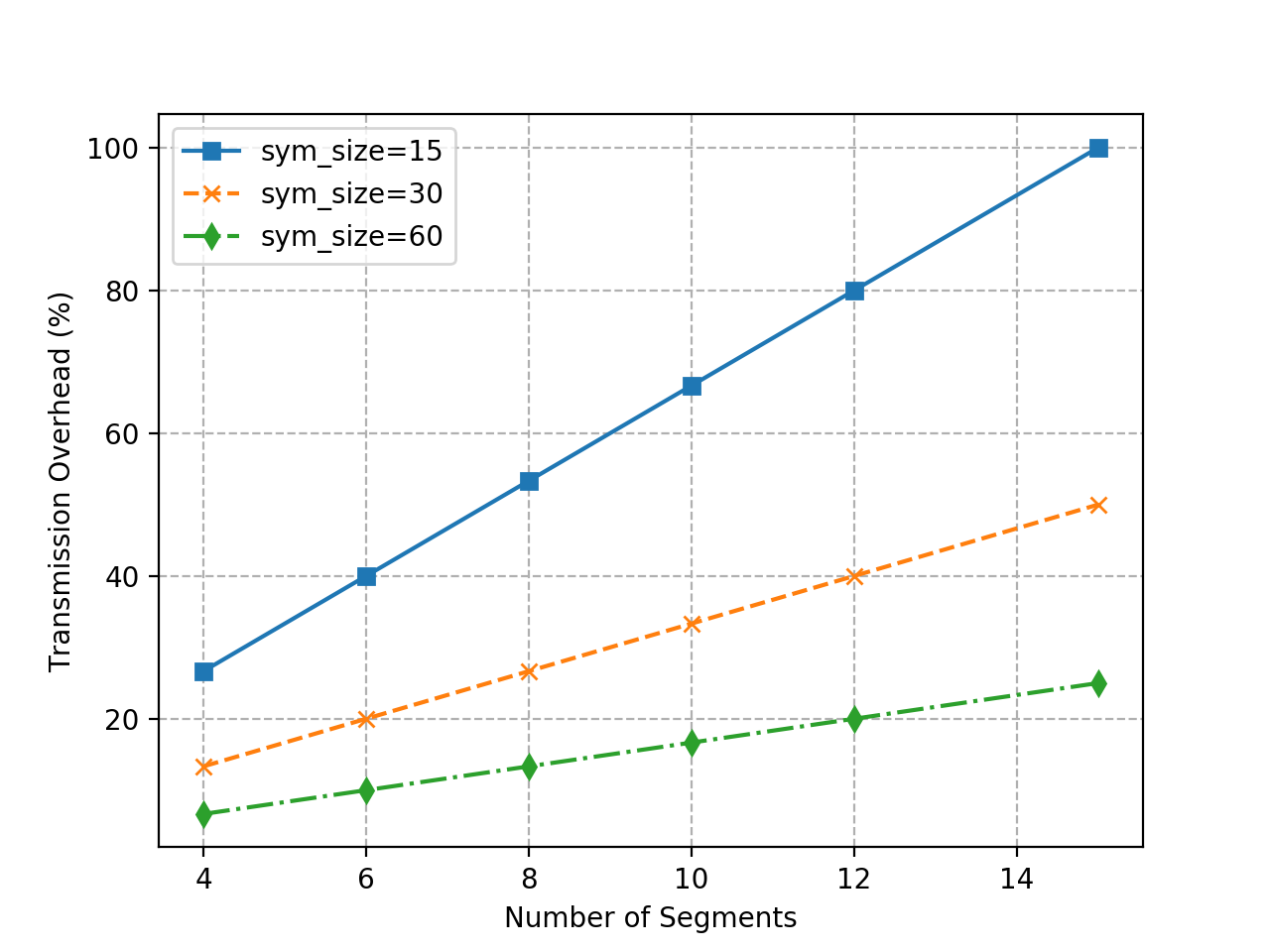}
	\caption{Overhead results of S-PRAC}
	\label{fig:overhead}
\end{figure}


\section{Evaluation} \label{sec:evaluation}

In this section, we describe our evaluation for S-PRAC and compare it with DAPRAC \cite{Cabrera2017}, the most notable related work. We describe the setup and metrics in Subsection \ref{subsec: implementation}. After that, we present and discuss the results in Subsection \ref{subsec: result&discussion}.


\subsection{Experimental Setup and Evaluation Metrics} \label{subsec: implementation}

We implement S-PRAC and DAPRAC using C++ and the fifi library \cite{fifi}.\footnote{~Our code can be found here: https://bitbucket.org/kirianto/sprac-cpp/src} We use two nodes, one as transmitter with encoder and one as receiver with decoder, and a point-to-point communication scheme. The transmission medium is very noisy, which causes many errors in the transmitted packets. 
Similar to \cite{Angelopoulos2014} and \cite{Cabrera2017}, we use RLNC for network coding. 

We experiment with different generation sizes (i.e. number of symbols), with a symbol size of 20 bytes. We first experiment with a small segment size of 8 bits. After that, we increase the segment size gradually to evaluate its impact on the performance. We repeated each experiment 10000 times. We plot below the average as well as the standard deviation.

In all the experiments, we selected a Galois field of two elements. A number of $g$ coded packets are created in the encoder with a systematic code fashion. Then, a varying number of corrupted bits are added to the first and the second coded packets. 
The receiver starts S-PRAC algorithm after receiving \emph{g}+1 packets.

We measure the performance of S-PRAC, and compare it to DAPRAC, using the following four metrics: (1) error estimation time (i.e. the time required to estimate locations of errors), (2) error correction time, (3) encoding time, and (4) decoding time which, as described in Subsection \ref{dec}, counts both the time required for running the scheme (S-PRAC or DAPRAC) as well as the time required for decoding the corrected packets. In addition, we calculate the overhead of S-PRAC by normalizing the total size of the inserted inner CRC-8 codes by the size of the original packet.

\subsection{Evaluation Results} \label{subsec: result&discussion}

\vspace{5pt}
\textbf{Error estimation time:} 
As shown in Fig. \ref{fig:err_estimation}, S-PRAC outperforms DAPRAC in terms of the time needed for estimating error locations, as long as the number of corrupted bits is low. However, this time increases in S-PRAC, and remains constant in DAPRAC, with the increase in the number of errors. These results are to be expected, and can be explained as follows: S-PRAC runs only on corrupted segments, which are likely few when the number errors is small, and high otherwise. In contrast, DAPRAC always checks the entire packet. 

\vspace{5pt}
\textbf{Error correction time:} 
Fig. \ref{fig:err_correction} plots the time required for correcting the errors. Here, S-PRAC remarkably  outperforms DAPRAC. The figure also shows that the more the errors, the larger the difference between S-PRAC results and DAPRAC results. For instance, when the number of errors is 10, S-PRAC could reduce the error correction time incurred by DAPRAC by more than 98\%. The superiority of S-PRAC is attributed to segmentation of packets, which results in fewer permutations, compared to DAPRAC (see Eq. \ref{eq: permu_prac} and Eq. \ref{eq: perm_sprac}). 

\vspace{5pt}
\textbf{Total of estimation time and correction time:} 
The superiority of S-PRAC is confirmed through the results plotted in Fig. \ref{fig:estimation_corr}. In particular, we can see that the total time required for both estimation and correction is always lower in S-PRAC than in DAPRAC. This superiority is more notable when the number of errors is high. For example, with 10 corrupted bits, the total time needed with S-PRAC is more than 90\% lower than the total time needed with DAPRAC. 

Fig. \ref{fig:segmentation} shows the impact of the number of segments on the total time needed for estimation and correction with S-PRAC. We can see that the more the segments, the lower the total time. This result is to be expected, because more segments translates into smaller segment sizes. 

\vspace{5pt}
\textbf{Encoding time and decoding time:} 
Fig. \ref{fig:encoding_decoding} complements the performance results discussed above. In particular, it shows the encoding and decoding times in S-PRAC and DAPRAC, under different generation sizes and two error values ($e=2$ errors and $e=10$ errors). We can see in Fig. \ref{fig:encoding} that the encoding, regardless of the number of errors, is slower in S-PRAC than in DAPRAC. This is because the encoding process in S-PRAC, in addition to encoding the packets using RLNC, includes two actions not existing in DAPRAC: (1) breaking down the packets into segments and (2) inserting the inner CRC codes. 

As for the decoding time, Fig. \ref{fig:decoding} shows that S-PRAC is slower when the number of errors is low ($e=2$), because the decoder in S-PRAC has to merge the segments back into a packet. However, DAPRAC becomes slower when we increase the errors ($e=10$). This lag is attributed to the additional permutations performed by DAPRAC in this case.  

\vspace{5pt}
\textbf{S-PRAC overhead:} 
Fig. \ref{fig:overhead} depicts the overhead results of S-PRAC, as defined in Subsection \ref{subsec: implementation}, for different numbers of segments and symbol sizes. We can see that the overhead, as to be expected, increases with the number of segments, and decreases with the symbol size. For instance, the overhead can be considered low (up to 20\%) as long as the packet is broken down into a maximum of six segments and, at the same time, the symbol size is not below 30. The overhead becomes above 50\% when the number of segments increases to 8 or more, while the symbol size is 15 or less. 

\section{Conclusion and Future Work} \label{sec:conclusion}

Our main contribution in this paper is S-PRAC, a fast solution for error detection and correction in very noisy wireless channels. The key idea of S-PRAC is to divide each packet into segments consisting of a certain number of RLNC-encoded symbols. S-PRAC also adds an inner CRC-8 code to each segment and an outer CRC-32 code to each coded  packet.

We evaluated S-PRAC through extensive experiments in a realistic setup. Overall, the results confirm the usefulness of S-PRAC. In particular, S-PRAC can remarkably lower the partial packet recovery time achieved with DAPRAC when the number of errors is high. This is achieved with a relatively low transmission overhead, unless the packet is broken down into a large number of segments (above 6).

For the future work, a further investigation and analysis of the overhead are needed for bandwidth optimization. Also, there are various existing error patterns based on channel conditions and characteristics in wireless networks. These have to be taken into consideration in future research. Moreover, the current design of S-PRAC can be improved in several ways. For instance, S-PRAC now cannot correct the errors when the inner CRC codes are corrupted, because the right permutation will never be achieved in this case. One solution to solve this problem is to calculate the outer CRC code only over the inner CRC codes, rather than on the whole packet. In addition, we are researching to improve S-PRAC performance in low noisy channels, and we also plan to implement and evaluate S-PRAC in real devices.
\section*{Acknowledgment}

This work is supported in part by the German Research Foundation (DFG), within the Collaborative Research Center SFB 912 -- HAEC.


\bibliographystyle{ieeetr}
\balance
\bibliography{bib/references}

\end{document}